\documentclass[11pt]{article}
\usepackage{graphicx}                  
\usepackage{amssymb}
\usepackage{amsmath}
\numberwithin{equation}{section}
\usepackage{multicol}
\usepackage{array}
\newcolumntype{C}[1]{>{\centering\arraybackslash}m{#1}}
\usepackage{epstopdf}
\usepackage{caption}
\usepackage{subcaption}
\usepackage{mathrsfs}
\usepackage{grffile}
\usepackage{floatrow}
\usepackage[amssymb]{SIunits}
\usepackage{floatflt} 
\usepackage[toc,page]{appendix}
\usepackage[listings,theorems]{tcolorbox}
\usepackage[section]{placeins}
\usepackage{wrapfig}
\usepackage{geometry}

\usepackage[utf8]{inputenc} 
\usepackage[amssymb]{SIunits}
\usepackage{geometry}
\usepackage{setspace}
\usepackage{physics}
\usepackage{authblk}
\usepackage[bookmarks]{hyperref}
\usepackage{titlesec}
\titleformat*{\section}{\large\bfseries}
\titleformat*{\subsection}{\normalfont\bfseries}
\titleformat*{\subsubsection}{\large\bfseries}

\title{\textbf{Taming the Delayed Choice Quantum Eraser}}

\author[1]{Johannes Fankhauser\thanks{\href{mailto:johannes.j.fankhauser@gmail.com}{\it johannes.j.fankhauser@gmail.com}}, \it University of Oxford.}

\date{(\today)}                                             

\begin{document}
\maketitle

\begin{abstract} 

I discuss the delayed choice quantum eraser experiment (DCQE) by drawing an analogy to a Bell-type measurement and giving a straightforward account in standard quantum mechanics. The delayed choice quantum eraser experiment turns out to resemble a Bell-type scenario in which the paradox's resolution is rather trivial, and so there really is no mystery.  At first glance, the experiment suggests that measurements on one part of an entangled photon pair (the idler) can be employed to control whether the measurement outcome of the other part of the photon pair (the signal) produces interference fringes at a screen after being sent through a double slit. Significantly, the choice whether there is interference or not can be made long after the signal photon encounters the screen. The results of the experiment have been alleged to invoke some sort of `backwards in time influence'. I argue that this issue can be eliminated by taking into proper account the role of the signal photon. Likewise, in the de Broglie-Bohm picture the particle's trajectories can be given a well-defined description at any instant of time during the experiment. Thus, it is again clear that there is no need to resort to any kind of `backwards in time influence'.

\end{abstract}


\newpage

\section{Introduction}

Delayed choice scenarios in slit experiments as found in \cite{wheeler1978past}, and earlier in \cite{von1941deutung} and \cite{bohr1996discussion}, have formed a rich area of theoretical and experimental research, as evidenced in the literature (\cite{eichmann1993young}, \cite{englert2000quantitative},  \cite{doi:10.1119/1.19257}, \cite{doi:10.1119/1.19258}, \cite{Kim1999}, \cite{walborn2002double}, \cite{kwiat2004science}, \cite{aharonov2005time},\\
\cite{Peres2000}, \cite{Egg2013}, to name a few). 
From the results of the original delayed choice experiment Wheeler concluded that  `no phenomenon is a phenomenon until it is an observed phenomenon', and `the past has
no existence except as it is recorded in the present' (ibid.). Others have also been inclined to conclude that such experiments entail some kind of backwards in time influence or another (e.g. \cite{doi:10.1119/1.19258}). I shall discuss a modified version of Wheeler's delayed choice experiment, which was first proposed by  Scully\textit{ et al.} in \cite{scully1982quantum} and later realised in the experiments of Kim \textit{et al.} in \cite{Kim1999}. It is termed the Delayed Choice Quantum Eraser. I show that the actual evolution of the quantum state in the experiment and a novel analysis in terms of a Bell-type scenario prove Wheeler's conclusions and other conclusions about backwards in time influence  unwarranted. The puzzlement about delayed choice experiments emerges from misinterpreting and ignoring the symmetry of time-ordered measurement events.  Since the analysis is general, it applies to all cases of quantum eraser where two systems become entangled. For a 3-slit quantum eraser experiment see, for example, \cite{shah2017quantum}.

\section{Delayed choice in a Bell-type scenario}
\label{Bell}

\begin{figure}[H]
\centering
\includegraphics[width=0.7\linewidth]{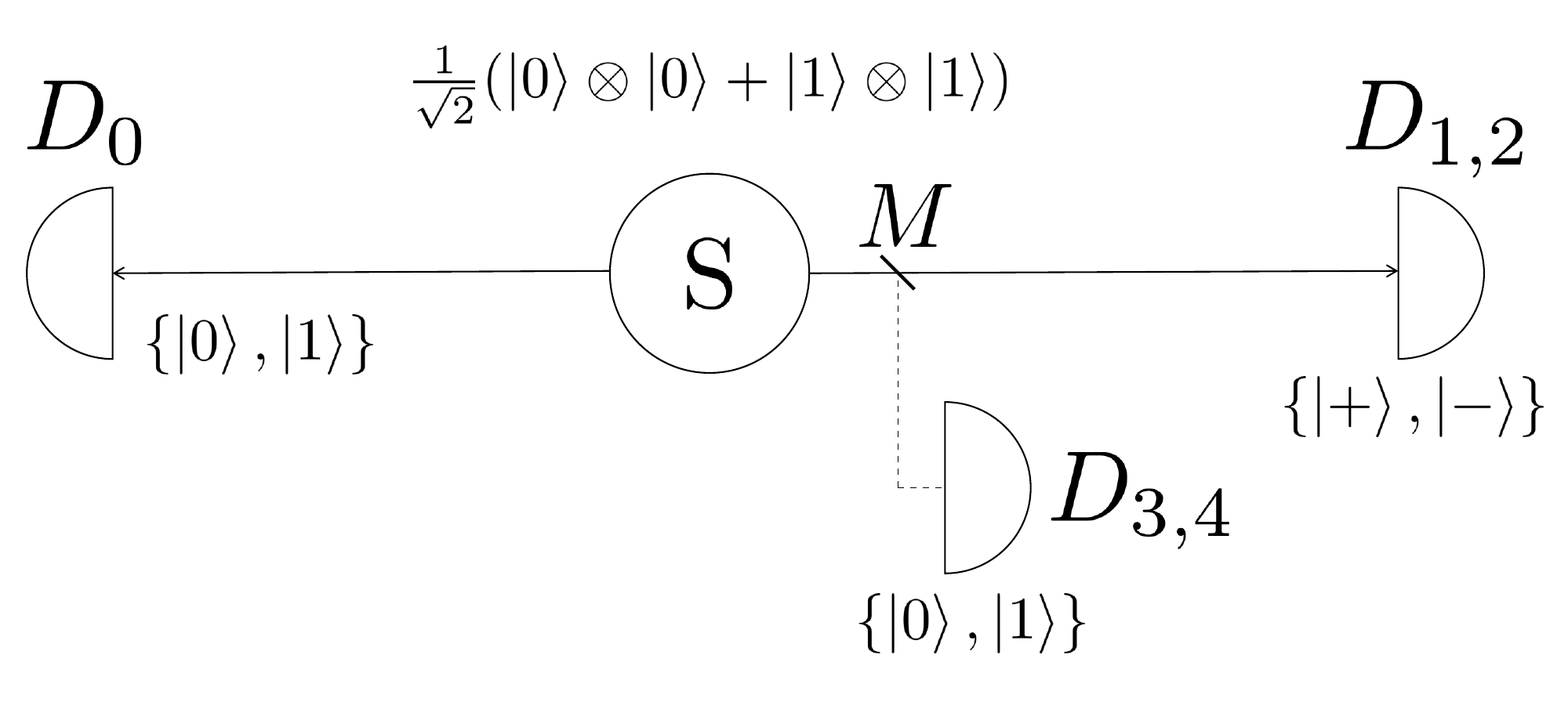}
\caption{A Bell-type experiment resembles the delayed choice quantum eraser experiment.}
\label{fig:BellD}
\end{figure}

Let us begin by considering a simple and familiar case, which will nonetheless provide the key to illuminating the DCQE. Imagine a source $S$ emitting photons. Both Alice (detector $D_0$) and Bob (detector $D_{1,2}$ and $D_{3,4}$) receive one particle of an entangled photon pair in the Bell state
\begin{equation}
\label{state}
\ket{\psi}=\frac{1}{\sqrt{2}}(\ket{0}\otimes\ket{0}+\ket{1}\otimes\ket{1}).
\end{equation} 
The states of the photons are taken to be qubit states. Let us, for ease of comparison with the DCQE later, call Alice's photon the signal photon and Bob's photon the idler. Figure \ref{fig:BellD} depicts the experiment.

$M$ denotes a mirror that can be used to reflect the idler photon into detector $D_{3,4}$. In Bob's arm two measurements can be performed: Either he chooses to use mirror $M$ to measure the idler photon in the computational basis $\{\ket{0},\ket{1}\}$ or the mirror is removed and the photon travels to detector $D_{1,2}$ where Bob performs a measurement in the diagonal basis  $\{\ket{+},\ket{-}\}$, where $\ket{+}=\frac{1}{\sqrt{2}}(\ket{0}+\ket{1})$ and $\ket{-}=\frac{1}{\sqrt{2}}(\ket{0}-\ket{1})$. 

The quantum predictions for the outcomes of the experiment are familiar and rather simple: The probabilities for the outcomes $0$ and $1$ at detector $D_0$ both are $0.5$ as easily seen from state $\ket{\psi}$ in Equation \ref{state}. And the same holds for the case in which Bob measures in the diagonal basis at detector $D_{1,2}$. For this latter case, the state is rewritten as   

\begin{equation}
\label{eraser state}
\ket{\psi}=\frac{1}{2}(\ket{0}\otimes(\ket{+}+\ket{-})+\ket{1}\otimes(\ket{+}-\ket{-})) =\frac{1}{2}((\ket{0}+\ket{1})\otimes\ket{+}+(\ket{0}-\ket{1})\otimes\ket{-}).
\end{equation}

The statistics at $D_0$ are independent of which measurement is performed by Bob, as expected. Furthermore, by conditioning on Bob's outcomes we can make the following statements:

(1) Assuming Bob's measurements come before Alice's (for, say, the rest frame of the laboratory) if he measured $\ket{0}$ at detector $D_{3,4}$, we know that Alice is going to measure $\ket{0}$ at $D_0$. Likewise, if the outcome was $\ket{1}$, she is going to measure $\ket{1}$. For the sake of comparison with the DCQE let us call this a `which-path measurement' since Bob's measurement will tell us with certainty which result Alice will subsequently observe. By contrast, if Bob decides to perform a measurement at detector $D_{12}$, conditioned on outcome $\ket{+}$ the state arriving at $D_0$ is going to be $\frac{1}{\sqrt{2}}(\ket{0}+\ket{1})=\ket{+}$. Thus, Alice will equally likely observe outcomes $0$ and $1$.  Conversely, conditioned on Bob measuring $\ket{-}$ Alice will receive the phase shifted state $\frac{1}{\sqrt{2}}(\ket{0}-\ket{1})=\ket{-}$ and she would again detect outcomes $0$ and $1$ with probability $0.5$. This directly follows from Equation \ref{eraser state}. We shall call this an `interference measurement' on Alice's side since a spread of results will be detected by Alice. 

(2) The case when Bob's measurements happens after Alice's is similar: When Alice obtains $0$, Bob is going to see $0$ as well in the $D_{3,4}$ measurement (the same holds for outcome $1$). In this case, just as previously, therefore, conditioning on Bob's recording a 0(1) outcome, Alice will have recorded a 0(1) outcome. In the case of the measurement at detector $D_{1,2}$, i.e. in the diagonal basis, Bob expects state $\frac{1}{\sqrt{2}}(\ket{+}+\ket{-}$ if Alice's outcome was $0$ and $\frac{1}{\sqrt{2}}(\ket{+}-\ket{+})$ otherwise. In both cases the outcomes $\ket{+}$ and $\ket{-}$ show up with probability $\frac{1}{2}$. Again, therefore, we see that the conditional statistics are the same as in the previous scenario; that is, the statistics are the same as  those which would be recorded if Bob's measurement had come first rather than Alice's: conditioned on Bob recording a +(--) result, Alice's will record 50\% outcome 0 and 50\% outcome 1.

It is important to note that on an operational view in terms of statistics of outcomes there is nothing puzzling about any of this, for there are no ontological commitments made other than the existence of conditional probabilities, which could just be understood as relative frequencies for certain pre- and post-selected subensembles. But as an exercise we could elicit a puzzle: We can now argue as follows that there must be retrocausal action. Bob is free to perform his measurements at any time. In particular, he can decide to perform a `which-path measurement' or 'interference measurement' well after the signal photon has reached Alice's detector. Since Bob by choosing to measure with detector $D_{1,2}$ or $D_{3,4}$ can decide to create either a computational basis state $\ket{0}$, $\ket{1}$ or a superposed state at Alice's detector ($\frac{1}{\sqrt{2}}(\ket{0}+\ket{1})$ or $\frac{1}{\sqrt{2}}(\ket{0}-\ket{1})$), the state of the signal photon that hit Alice's detector had to change retroactively in order to get the outcomes expected. In other words, Bob's measurement determines the state of the signal photon, but since that photon has already been measured, it must have done so by acting on the past of it. Moreover, we might argue that one `has' to reason in this way, since the statistics that we obtain for Alice's outcomes when we condition on Bob's later outcome are `exactly those' which are generated when Alice receives a quantum state produced as a result of Bob's measurement.

The error in this naive argument for `action into the past' applied to the Bell-type scenario, is apparent immediately. It is true that conditioned on Bob's outcome (if it happens first) we can infer the signal photon's state, but if Alice's measurement happens first we need to condition Bob's state as well! Thus, story (2) is to be told. In a nutshell, the puzzle arises from ignoring the role of the photon that hits detector $D_0$ conditioned on whose outcome explains the behaviour at Bob's site. If the outcome of measuring the idler at $D_{1,2}$ is, say, $\ket{+}$, would we expect that the measurement to have changed the past of the other particle to $\frac{1}{\sqrt{2}}(\ket{0}+\ket{1})$? Certainly not. Only when the signal photon has not yet encountered detector $D_0$ would we say it evolved to $\frac{1}{\sqrt{2}}(\ket{0}+\ket{1})$ given that the state of the idler photon yielded $\ket{+}$. Otherwise, the outcome at of Alice's measurement will first give $\ket{0}$ or $\ket{1}$ and, as a result, leave the state of the idler photon in a mixed state of $\ket{+}$ and $\ket{-}$.

Two comments are in order. For one, the puzzle only has any grip since the outcome statistics of Alice measuring first and Bob after equals the outcome statistics of Bob measuring first and Alice measuring after. Obviously, the time order of which measurement happens first does not matter as is clear from the wavefunction of the system and is anyway enforced by the no-signalling theorem of quantum mechanics. 
This allows the post-selection to be done after the outcomes have occurred. Violation of this condition would indeed lead to a genuine paradox, and would in fact allow for signalling. Noticing this symmetry does not dispel the paradox but is the reason for why people can get confused about what is going on in the experiment. 

For the other, assuming the signal photon to be in one of the basis states before it or the idler photon were measured puts one into the business of hidden variable theories (after all, non-locality has it that the actual state of Alice indeed changes when Bob performs his measurement; see Section \ref{Bohm}). For this introduces an ontological commitment as to the value definiteness of states prior to measurement. For example, both Egg and Ellerman correctly point out that if a detector can only detect one collapsed eigenstate this does not mean that the photon was already in that state prior to that measurement (\cite{Egg2013,ellerman2011very,Ellerman2015}). That's why one might want to avoid  phrases like `which-path information' as there isn't any information about which-path since no path was ever uniquely taken. 
Most importantly, one can tell a coherent story about the states in the experiment without resorting to `retrocausal action into the past'. In lights of this analysis the puzzle seems trivial.\footnote{It was drawn to my attention that a somewhat similar conclusion was very recently independently reached in \cite{kastner2019delayed}.} 

In the double slit DCQE experiment (Section \ref{Scully}) the paradox is more disguised by the details of the experiment, but we will see that the same story can be told as in the Bell scenario.

\section{The Double Slit Delayed Choice Quantum Eraser}
\label{Scully}
The setup employed by Kim et al. uses double slit interference of photons and raises a conceptual problem, which, according to Wheeler and others would allegedly imply that there was a change in the behaviour from `acting like a particle' to `acting like a wave', or vice versa, well after the particle entered the double slit.  

\begin{figure}[H]
\centering
\includegraphics[width=0.8\linewidth]{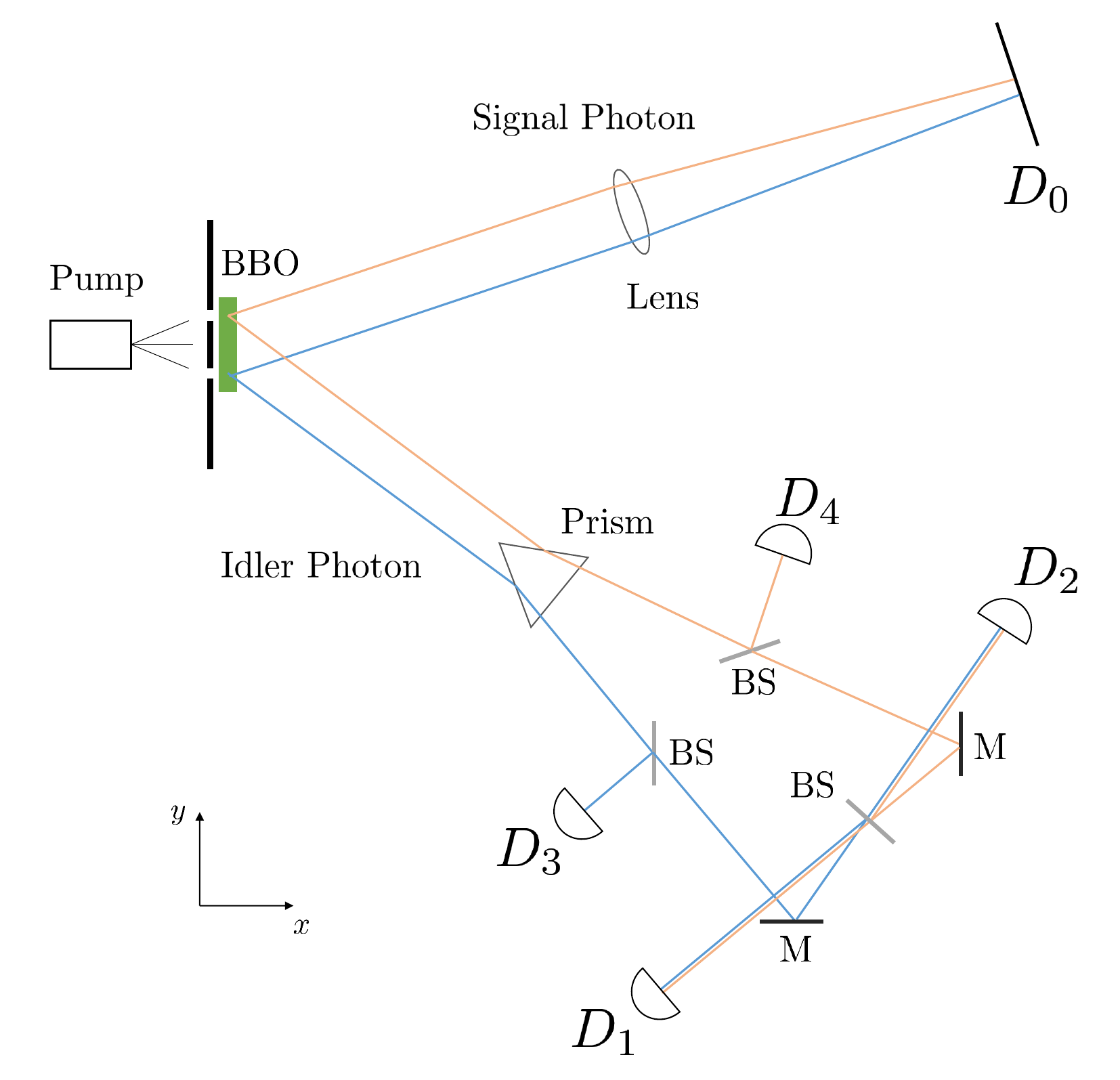}
\caption[]{A delayed choice quantum eraser experiment. A laser beam aims photons at a double slit. After a photon passes the slits it impinges on a Barium borate (BBO) crystal placed behind the double slit. The optical crystal destroys the incoming photon and creates an entangled pair of photons  via spontaneous parametric down conversion at the spot where it hit. Thus, if one of the photons of the entangled pair can later be identified by which slit it went through, one also knows whether its entangled counterpart went through the one or the other side of the crystal. Whether which-path information about the signal photon arriving at detector $D_0$ is obtained or erased is decided by manipulating the idler photon well after the signal photon has been registered.}
\label{fig:Kim}
\end{figure}

In the old days of quantum mechanics it was believed that the loss of interference in double slit experiments were due to Heisenberg's uncertainty principle, for no measurement device could be so fancy as not to perturb the system observed and destroy coherence. Such a perturbation leads to so-called `which-path information' that `collapses the wavefunction', making interference effects disappear. More concretely, when an interaction of the quantum particle with the measurement device occurs, the terms in the quantum state that led to interference are coupled with states of the measurement device, and those states of the combined system are orthogonal for a reasonable measurement device. In the DCQE case the which-path information of the photon is obtained by entanglement with an auxiliary photon without disturbing the wavefunction (cf.~Einstein's move in the EPR experiment \cite[p. 779]{PhysRev.47.777}). Significantly, the which-path information can be `erased' long after the photon encounters the double slit. This is possible by further measurement procedures on the entangled photon. The interference pattern, as a result, reappears.  This was deemed inconceivable in the old picture since as soon as a measurement happen the state was believed to have been irrevocably disturbed. Figure \ref{fig:Kim} illustrates the experimental setup. 
The usual story of how the DCQE works goes like this: A laser beam (pump) aims photons at a double slit. After a photon passes the slits it impinges on a Barium borate (BBO) crystal placed behind the double slit. The optical crystal destroys the incoming photon and creates an entangled pair of photons  via spontaneous parametric down conversion at the spot where it hit. Thus, if one of the photons of the entangled pair can later be identified by which slit it went through, we will also know whether its entangled counterpart went through the one or the other side of the crystal.   
By contrast, we will have no which-path information if we cannot later identify where either of the photons came from. Even though the entangled photons created at the crystal are now correlated, the experiment can manipulate them differently. We call one photon of the pair the signal photon (sent toward detector $D_0$) and the other one the idler photon (sent toward the prism). The naming is a matter of convention. The lens in front of detector $D_0$ is inserted to achieve the far-field limit at the detector and at the same time keep the distance small between slits and detector. The prism helps to increase the displacement between paths. Nothing about these parts gives which-path information and detector $D_0$ can not be used to distinguish between a photon coming from one slit or the other. At this point we might expect interference fringes to appear at $D_0$ if we were to ignore that the signal photon and idler photon are entangled. Considering the signal photon alone we might think that the parts of the wavefunction originating at either slit should interfere and produce the well-known pattern of a double slit experiment. On the other hand, quantum mechanics would predict a typical clump pattern if by taking into account the idler photon, which-path information were available.
 
After the prism has bent the idler photon's path, the particle heads off to one of the 50-50 beamsplitters \textit{BS}. The photon is reflected into the detector $D_3$ a random 50\% of the time when it is travelling on the lower path, or reflected into detector $D_4$ a random 50\% of the time when it is travelling on the upper path. If one of the detectors $D_3$ or $D_4$ clicks, a photon is detected with which-path information. That is, we know at which slit both photons of the entangled pair were generated. In that case, the formalism of quantum mechanics predicts no interference at $D_0$. In all of the other cases the photon passes through the beamsplitter and continues toward one of the mirrors $M$. Importantly, it does not matter if the choice whether the photon is reflected into the which-path detectors $D_3$ or $D_4$ is made by beamsplitters. The original experiment uses beamsplitters and therefore it is randomly decided which kind of measurement is performed. But we could equally replace the beamsplitters by moveable mirrors. In that way the experimenter is free to decide whether which-path information is available by either keeping the mirrors in place or removing them such that the photon can reach the eraser. 

After being reflected at one of the mirrors, the photon encounters another beamsplitter $BS$, which is the quantum eraser. This beamsplitter brings the photon into a superposition of being reflected and transmitted. To that end, for an idler photon coming from the lower mirror the beamsplitter either transmits the photon into detector $D_2$ or reflects it into detector $D_1$. Likewise, for an idler photon coming from the upper mirror the beamsplitter either transmits it into detector $D_1$ or reflects it into detector $D_2$. If one of the detectors $D_1$ or $D_2$ clicks, it is impossible to tell which slit the photon came from. 
To summarise the above, detectors $D_1$ and $D_2$ placed at the output of $BS$ erase the which-path information, whereas a click of detectors $D_3$ or $D_4$ provides which-path information about both the idler and the signal photon. 
Notably, when the photon initially hits $D_0$, there is no which-path information available, only later when the entangled idler photon is detected at $D_3$ or $D_4$.
\begin{figure}[H]
\centering
\includegraphics[width=1\linewidth]{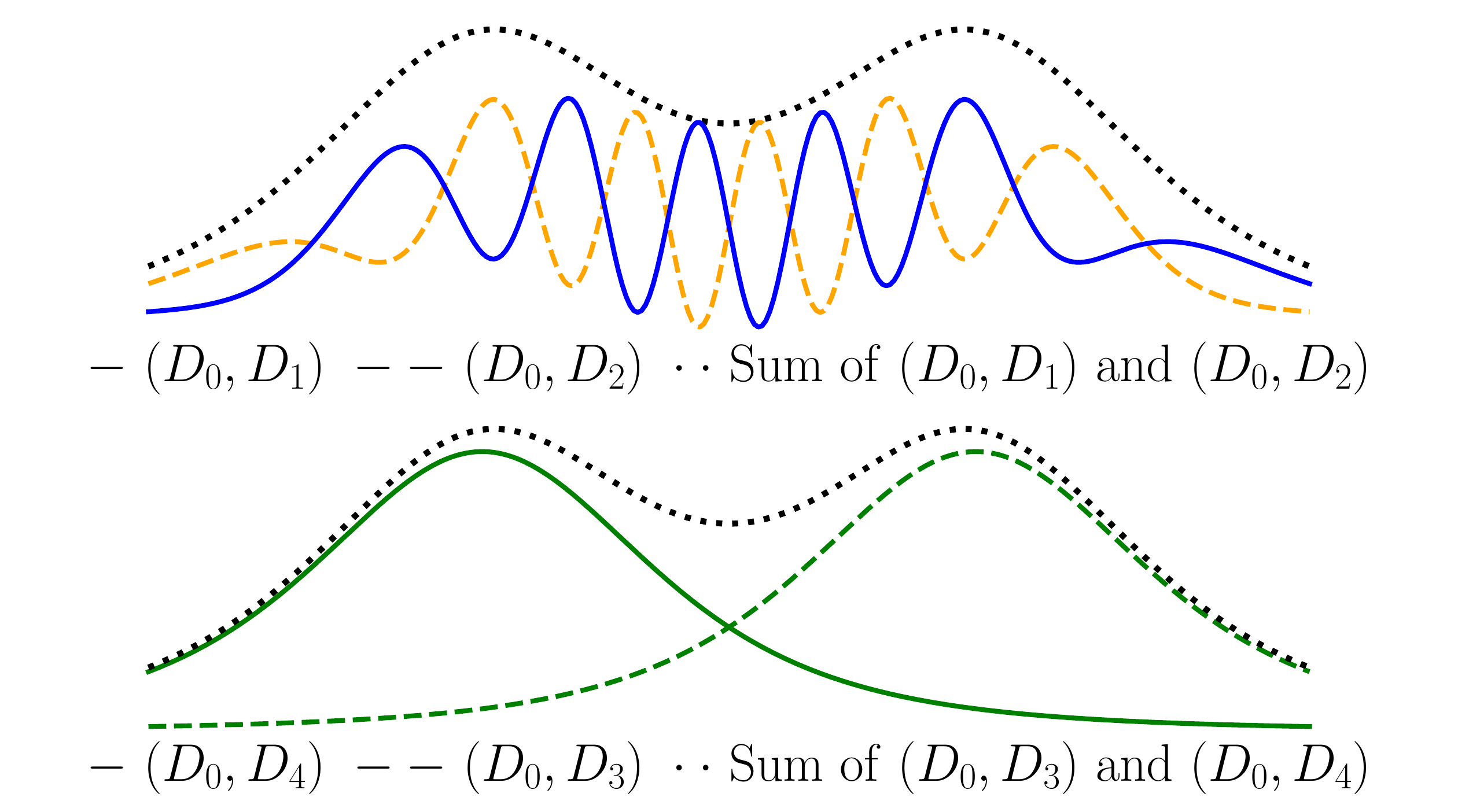}
\caption{Joint detection events at detector $D_0$ and detectors $D_1$-$D_4$. The figure shows a plot of the bits of Equation \ref{expression}. Records of $D_0$ and $D_1$ ($D_2$) show interference fringes. Conversely, records of $D_0$ and $D_3$ ($D_4$) show a clump pattern. If added together, the distribution on the screen in both cases (with and without eraser) is always two clumps.}
\label{fig:patterns}
\end{figure} 
This is key. The setup ensures that the which-path information is only erased or provided, respectively, after $D_0$ has detected the signal photon. We therefore say the choice is delayed.  
For each incoming photon from the laser beam there will be a joint detection of the signal photon at $D_0$ and the idler photon at $D_1$--$D_4$. Figure \ref{fig:patterns} shows the expected results.\footnote{The results in \cite{Kim1999} show a single clump as opposed to two clumps in Figure \ref{fig:patterns}. This is simply due to the close distance between the slits Kim et al. chose for their experiments.} When which-path information is provided, a clump pattern appears, but when no which-path information is available interference fringes appear. The two interference patterns corresponding to correlation with $D_1$ and $D_2$ are out of phase. The reason for that will become clear in the next sections. 

Those inclined to instrumentalism might be satisfied at this point, for the predictions of standard quantum mechanics give the desired results to confirm experimental observation. The philosopher, however, might start to worry about what is going on here. 

\section{Backwards in time influence?}
Indeed, it may be tempting to interpret these results as instances of future measurements influencing past events. Seemingly, there is something odd going on in the experiment. The appearance on the screen (either one that shows interference or one that shows a clump pattern) of the signal photon is determined by the way in which the idler photon is measured --- a choice that occurs after the signal photon has already been detected. Can a later, distant measurement cause an entangled particle to retroactively alter its wavefunction? It seems the detection of the idler photon and thus the choice of which-path information affects the behaviour of the signal photon in the past. Is this a process that reverses causality? Wheeler comments on his original \textit{Gedankenexperiment} as follows:
\begin{quote}
\textit{`Does this result mean that present choice influences past dynamics, in contravention of every formulation of causality? Or does it mean, calculate pedantically and do not ask questions? Neither; the lesson presents itself rather like this, that the past has no existence except as it is recorded in the present.'} \cite{wheeler1978past}
\end{quote}
Thus, Wheeler accepts that the past could be created \textit{a posteriori} by happenings in the future. In comparison, Bohr concludes that understanding of the quantum behaviour of particles is confused by giving pictures which are trying to maintain conceptions of classical physics. He states that a sharp separation of the quantum system and the observing measurement device is impossible \cite{bohr1961atomic}. According to his view there is no point in visualising the process as a path taken by a particle when not in a well-defined state.  In a more recent work Brian Greene argues that delayed choice quantum eraser scenarios may not alter the past but future measurements certainly determine the story we tell about the past behaviour of a particle. His account, although, is too vage to reach a satisfactory resolution.  \cite{brian2004fabric}[pp. 194-199] 

One should not expect the formalism of quantum mechanics to provide clear images of what could be actually going on, for at the moment it is a framework with different interpretations. Only if one is to adopt an interpretation, I believe, can a conclusion be meaningful. Many physicists and philosophers did not accept the views of Wheeler or Bohr and have been continuing to debate the delayed choice experiment to seek for possibilities that account for physical intuition. 

\section{Delayed choice in standard quantum mechanics}

The first significant point is that there never appears an interference pattern at $D_0$ without conditioning on whether we choose which-path information to be available or erased.\footnote{Note that in the experiment of Kim et al. the decision is made randomly by the beamsplitters next to the prism, but as I mentioned, they can be replaced with mirrors and allow the experimenter to make this choice.} Technically, by conditioning we mean to constrain the already observed measurement results to the subset of coincidence detections of the signal photon with the idler photon in a chosen detector $D_1$--$D_4$.  Moreover,  the two interference patterns from the joint detection events of $D_0$ and $D_1$ or $D_2$, respectively, obtain a relative phase shift of $\pi$ and cancel when added together. This feature is often left out but crucial as we shall see.    

I shall give an analysis of the experiment proposed by Kim \textit{et al.} by using standard quantum mechanics.\footnote{Egg discussed Scully's delayed choice version involving a cavity to distinguish between which-path measurement and interference measurements \cite{Egg2013}. He gives an account of the experiment in standard quantum mechanics. Although he presents some insights on metaphysical claims on entanglement realism that are taken into question by delayed choice scenarios. But, in my view, there is too little emphasis on the importance of the point that the paradox results from misinterpreting the pre and post measurement symmetry in quantum mechanics and the significance of detecting the signal photon and condition on its outcome. Having said that, it might not have been Egg's endeavour to resolve the paradox in the DCQE in the way I am describing here since I think the main point for him was to argue against non-realist arguments (e.g. in \cite{healey2012quantum}) that seem to undermine the physical reality of entanglement by means of delayed choice scenarios.} The wavefunctions involved are described by Schrödinger's equation, which strictly speaking only applies to massive particles. For a rigorous treatment with photons we would need to avail ourselves of quantum field theory. Nevertheless, we can straightforwardly replace photons with electrons for the sake of a \textit{Gedankenexperiment}. The interference phenomena qualitatively remain the same. I shall retain the term `photons' in the derivation throughout the paper for clarity, however. 

The incoming laser beam can be described as a plain wave 
\begin{equation}
\label{laser}
\psi=e^{ik_xx}
\end{equation} impinging on the double slit, where $k_x$ is the wave vector.\footnote{For the sake of simplicity we can suppress time dependence of the wavefunction since it does not affect the argument. I omit normalisation factors where not stated explicitly.} After the slits the wavefunction can be decomposed into two interfering parts as
\begin{equation}
\psi= \frac{1}{\sqrt{2}}(\psi_1+\psi_2).
\end{equation} Wavefunction $\psi_1$ belongs to the part of the wavefunction emerging from the upper slit and $\psi_2$ to the part of the wavefunction emerging from the lower slit. We may assume waves of the form
\begin{equation}
\psi_i=\frac{e^{ikr_i}}{r_i},
\end{equation} where $r_i$ is the distance from the slit $i$. 
These give the well-known two slit interference fringes. The crystal then creates an entangled pair of photons with opposite momenta in the $y$-direction such that 

\begin{equation}
\label{pair}
\psi=\frac{1}{\sqrt{2}}(\psi_1\otimes \psi_1'+\psi_2\otimes \psi_2'),
\end{equation} where unprimed wavefunctions correspond to the signal photon and primed to the idler photon. The signal photon sent to detector $D_0$ is now entangled with the idler photon. This affects the probability amplitudes at $D_0$, and interference between $\psi_1$ and $\psi_2$ vanishes since $\psi_1\otimes \psi_1'$ and  $\psi_2\otimes \psi_2'$ are orthogonal states. Note that $\psi_1'$ and $\psi_2'$ are thought to be non-overlapping and thereby the inner product vanishes. Note also that in general orthogonality of two states does not imply zero overlap of the states in position basis. The squared norm of the wavefunction yields

\begin{equation}
|\psi|^2= \frac{1}{2}(|\psi_1|^2|\psi_1'|^2+|\psi_2|^2|\psi_2'|^2).
\end{equation} By integrating out the idler degrees of freedom we find for the probability distribution of the signal photon on the screen
\begin{equation}
\rho= \frac{1}{2}(|\psi_1|^2+|\psi_2|^2).
\end{equation} From this it is clear that no interference will appear on the screen for this state. Assuming the signal has not yet reached $D_0$, if the idler gets reflected into detector $D_3$ the state would be $\psi_2\otimes \psi_2'$, and if reflected into $D_4$ it would be $\psi_1\otimes \psi_1'$. 
In case the idler photon encounters the quantum eraser, the wavefunction undergoes another unitary evolution. The eraser puts the idler photon in a superposition of being transmitted to one detector or reflected to the other. At each reflection at a beamsplitter or mirror the wavefunction picks up a phase of $\frac{\pi}{2}$ (a multiplication of the wavefunction by $e^{i\frac{\pi}{2}}=i$) such that
\begin{align}
\psi_1' &\mapsto \ ~ i\psi_{D_1}-\psi_{D_2}\nonumber\\
\psi_2' &\mapsto  -\psi_{D_1}+i\psi_{D_2}.
\end{align} The joint wavefunction then turns into
\begin{align}
\label{expression}
\psi &= \frac{1}{2}(\psi_1\otimes (i\psi_{D_1}-\psi_{D_2})+\psi_2\otimes (-\psi_{D_1}+i\psi_{D_2})) \nonumber \\ &=\frac{1}{2}((i\psi_{1}-\psi_{2})\otimes\psi_{D_1} + (-\psi_{1}+i\psi_{2})\otimes\psi_{D_2})
\end{align} 
once the idler photon has passed the quantum eraser. Indices in $\psi_{D_1}$, $\psi_{D_2}$ refer to which detector the part of the wavefunction is reflected into. In this form state \ref{expression} makes it clear that when detector $D_1$ clicks, conditioned on this event the state of the signal photon is $i\psi_{1}-\psi_{2}$, yielding a probability distribution of interference fringes, 
\begin{align}
|\psi_{D_0, D_1}|^2 &= (i\psi_{1}-\psi_{2})\overline{(i\psi_{1}-\psi_{2})}\nonumber \\
&=|\psi_1|^2+|\psi_2|^2-2\Im(\overline{\psi_1}\psi_2).
\end{align} In the case in which $D_2$ clicks, conditioned on that event the state is $-\psi_{1}+i\psi_{2}$ and yields a distribution showing shifted anti-fringes:
\begin{align}
|\psi_{D_0, D_2}|^2 &= (-\psi_{1}+i\psi_{2})\overline{(-\psi_{1}+i\psi_{2})} \nonumber 
\\ &=|\psi_1|^2+|\psi_2|^2-2\Im(\psi_1\overline{\psi_2}) \nonumber
\\ &=|\psi_1|^2+|\psi_2|^2+2\Im(\overline{\psi_1}\psi_2).
\end{align} In both of the cases, there is a path on which the idler is reflected twice, and a path on which it is reflected once.

So far there is no puzzle. The experiment of Kim \textit{et al.}, however, is designed such that the choice whether the state produces interference fringes or a clump pattern happens after the signal photon has been detected at $D_0$. We therefore say the choice is delayed. In the setup of \cite{Kim1999} the optical length of the idler photon is about $8\ \nano\second$ longer than that of the signal photon. If we accepted that the causal story to be told about what is going on with the signal photon at any time is purely determined by what happens to the idler photon, then this would suggest backwards action from the future since the measurement of the signal photon has already occurred. With all this in mind, must we conclude that a measurement in the present retroactively changes the past to make it agree with the measurement outcomes? 

Crucially, at detector $D_0$ there never appears an interference pattern, regardless of whether the idler photon reaches the quantum eraser or not. This can readily be seen by adding up the distributions:
\begin{equation}
|\psi_{D_0, D_1}|^2 + |\psi_{D_0, D_2}|^2 = |\psi_1|^2+|\psi_2|^2.
\end{equation} The interference terms cancel out when added together which effectively leads to a clump pattern. Each sub-case shows an interference pattern, but the overall statistics adds up to two clumps. Note that there is no way to avoid the phase difference in the interference fringes since any additional device would act symmetrically on both paths. Insert for instance a $\lambda/4$-plate into the paths of the idler photon and it will affect both of the superposed paths reflected into the detectors. Thus, the effect of the plate would cancel out.

Of course, the fact that at detector $D_0$ interference fringes never occur guarantees consistency with no-signalling between $D_0$ and the other detectors. That is to say, it is not possible to decide what distribution (either an interference pattern or a clump pattern) appears at the detector $D_0$ by choice of whether the idler photon will trigger the which-path detectors $D_3$ and $D_4$ and thus communicate information.  As I noted above, this choice can be realised by replacing the former two beamsplitters by mirrors which can be inserted \textit{ad libitum} by the experimenter (compare no-signalling in \textit{EPR}).

The apparent retroactive action vanishes if a click in $D_0$ is regarded to condition the overall wavefunction too, not only a click in the detectors $D_1$--$D_4$ (think back to the discussion of the Bell-type experiment in Section \ref{Bell}). In the standard explanation, if the detection of the idler photon happens before the detection of the signal photon at $D_0$, the detectors $D_1$--$D_4$ determine what state the signal photons end up in. But similarly, in the case when the signal photon is detected at a moment in time preceding the observation of the idler photon, the detected position of the signal photon determines the state of the idler photons which will go on to trigger one of the detectors $D_1$--$D_4$. 

We can see this by rewriting state \ref{pair} in the position basis of the signal photon. Let's first work out the state for the which-path measurement:
\begin{align}
\psi&=\frac{1}{\sqrt{2}}\int(\psi_1(x)\ket{x}\otimes \ket{\psi_1'}+\psi_2(x)\ket{x}\otimes \ket{\psi_2'})dx\\ \nonumber
&= \frac{1}{\sqrt{2}}\int\ket{x}\otimes(\psi_1(x) \ket{\psi_1'}+\psi_2(x)\ket{\psi_2'})dx
\end{align}
From this we can see that if the signal photon hits the screen at position $x$ the probability for a click in $D_3$ is (roughly, since the two lumps slightly overlap) $|\psi_2(x)|^2$. And for detector $D_4$ it is roughly $|\psi_1(x)|^2$.  In other words, we can be almost sure which of the two detectors $D_3$ or $D_4$ will fire by looking at what lump on the screen the signal photon ends up in. Note that, vice versa, conditioned on a click in the respective detectors we previously found  $\psi_2(x)$ or $\psi_1(x)$ to be the state determining the distribution of signal photon hits on the screen. Thus, the two causal stories are consistent and the same correlations between signal and idler photon arise as expected. Note that for a quantum eraser the wavepackets of the signal photon need to overlap sufficiently in position space in order to give rise to interference that can be `erased'. This is true at the screen and in the overlap reagion before the screen. Where the wavepackets do not overlap, there is a set of quantum particles that do not interfere (compare also with \cite{Quanta87}). In the idealised case of circular waves considered here, the waves in fact overlap at all points in space between source and screen. 

Conversely, by rewriting Equation \ref{expression} for the interference measurement after the idler passed the quantum eraser we obtain
\begin{align}
\psi &= \frac{1}{2}\int[\psi_1(x)\ket{x}\otimes (i\ket{\psi_{D1}}-\ket{\psi_{D2}})+\psi_2(x)\ket{x}\otimes (-\ket{\psi_{D1}}+i\ket{\psi_{D2}})]dx \nonumber \\ &=\frac{1}{2}\int\ket{x}\otimes[(i\psi_1(x)-\psi_{2}(x))\ket{\psi_{D1}}+(-\psi_{1}(x)+i\psi_{2}(x))\ket{\psi_{D2}}]dx.
\end{align} The probability for $D_1$ to fire is $|i\psi_1(x)-\psi_{2}(x)|^2$ and for $D_2$ it is $|-\psi_1(x)+i\psi_{2}(x)|^2$ if the signal photon was detected at position $x$ on the screen. Again, as before, we recover that the probabilities are consistent with the time-reversed story where the idler photon is detected first. That is, state $i\psi_1(x)-\psi_{2}(x)$ determines the outcomes on the screen for conditioning on $D_1$ and state $-\psi_1(x)+i\psi_{2}(x)$ for conditioning on $D_2$.


One faces a confusion if one is to stubbornly stick to the notion that a measurement of the idler photon determines the probability distribution at $D_0$ for the signal photon. In fact, observation of individual subsystems of entangled pairs never changes the probability distribution of the remote particle. After all, the conditional probabilities of the measurement outcomes of signal and idler photon are spatio-temporally symmetric. Gaasbeek tried to reason in \cite{Gaasbeek2010} that this idea alone were to demystify the paradox. Though, I wish to emphasise again that the symmetry in the time-ordering is crucial to realising why the alleged paradox arose in the first place. If the two causal stories I just outlined above gave \textit{different} predictions on the probabilities of outcomes depending on which photon is detected first, it would be clear that there cannot be backwards in time influence since from the patterns on the screen one could tell which measurement happened first. However, as the probabilities are invariant under the time-order of measurements on the signal and idler photon, one can get confused as to whether the idler photon could retroactively determine the patterns on the screen. 


What this tells us is that no matter how the idler photon gets manipulated, the probability distribution on $D_0$ is a clump pattern, but when we condition on the outcome of the detectors, which either give which-path information or not, we find correlations as expected and, most importantly, the same correlations arise when the conditioning on the outcome of the signal photon. The quantum eraser does not influence the past of the signal photon; rather it reveals the correlations of an entangled photon pair in just another way. This only is puzzling because the probabilities conditioned on the postselection are time symmetric in the pre and post selected states. 

To reinforce the point compare the situation with the Bell scenario in Section \ref{Bell}: The source $S$ of an entangled pair of photons can be identified with the laser beam, the double slit, and the BBO crystal. $M$ denotes a mirror that can be used to reflect the idler photon into $D_{3,4}$. In the Bell scenario detectors $D_3$ and $D_4$ were concatenated into one detector, where an outcome $\ket{0}$ would correspond to detection at $D_3$ and an outcome $\ket{1}$ to detection at $D_4$. We stipulate that the signal photon is sent towards the lens and the idler photon to the prism. If we are to perform a which-path experiment we measure the idler photon in the computational basis $\{\ket{0},\ket{1}\}$ at $D_{3,4}$. Detector $D_0$ measures the signal photon in the computational basis, which corresponds to an interference measurement if the state of the signal photon, for instance, is one of the states of the diagonal basis $\{\ket{+},\ket{-}\}$. The measurement on the idler photon in the diagonal basis (at $D_{1,2}$) acts as the quantum eraser, i.e. a measurement of the idler photon in the diagonal basis is consistent with the signal photon being in a supersposition of $\ket{0}$ and $\ket{1}$. The results of the detectors $D_0$ conditioned on the outcome of $D_{1,2}$ show the familiar correlations when compared. 

When a photon has past the BBO crystal, the quantum state ends up in an entangled one. After rewriting the second slot of the state in the diagonal basis, we recover a wavefunction that is qualitatively identical to Equation \ref{expression}. Thus, the paradox in the double slit case resolves in the same ways as the one we alleged to the Bell-type scenario in Section \ref{Bell}.


\section{Delayed choice in de Broglie–Bohm theory}
\label{Bohm}

In Bohmian mechanics or de Broglie-Bohm's theory particles follow definite trajectories at all times. Thus, working out the DCQE in such a framework seems particularly appealing when it comes to verifying whether past trajectories could in any way depend on future measurements. As it turns out, as well as in the standard quantum treatment, in the hidden variable approach the puzzle resolves. Although, for instance, Hiley and Callaghan treated a double slit version of the delayed choice quantum eraser in Bohmian mechanics in \cite{hiley2006erased}, it is instructive to treat the DCQE version described here. Hiley and Callaghan analyse a case in which which-path information is acquired by a cavity wherein atoms get excited. This leads to interference fringes and anti fringes that do not appear in the setup employed by Kim \textit{et al.} Therefore, I shall work out the ongoings of Kim's setup within de Broglie-Bohm's theory too in the following. 

I will use the term `de Broglie-Bohm theory' to stand for the interpretation discussed by \cite{bohm2006undivided}. Here it is assumed that a particle always travels on only one path. The wavefunction is considered as a quantum potential or pilot wave and used in its polar form

\begin{equation}
\psi(\vec{r},t)=R(\vec{r},t)e^{iS(\vec{r},t)/\hbar}.
\end{equation} The dynamics of the pilot wave obey the Schrödinger equation
\begin{equation}
i\hbar\partial_t\psi=H\psi
\end{equation} and the particle's trajectory is determined by

\begin{equation}
\vec{v}\left(t\right)=\dot{\vec{x}}(t)=\frac{1}{m}\nabla S(\vec{r},t)|_{\vec{r}=\vec{x}}
\end{equation} where $m$ is the mass of the particle. For the sake of simplicity I will set $\hbar=1$ for the remainder.

Now let us turn to consider how particles behave according to de Broglie-Bohm in this experiment. We construct a set of possible trajectories, each individually corresponding to one initial value of position of the particle within the incident beam. Supposedly, de Broglie-Bohm theory should reveal whether the past is influenced by present observations since it assumes a well-defined path of the particles at all times. Note that the de Broglie-Bohm interpretation does allow us to illustrate such a process and reproduce all the known experimental results in tension with Wheeler's and Bohr's conclusion about these phenomena.

The wavefunction of the incoming laser beam \ref{laser} is already in polar form and the trajectories in this region are straight lines. First we consider the case without the eraser. To work out what happens we must write the final wavefunction in Equation \ref{pair} in the form\footnote{For simplicity I suppress normalisation factors.}

\begin{equation}
\psi(r,r')= R(r,r')e^{iS(r,r')}.
\end{equation} The wavefunction is evaluated at the positions of the signal photon $r$ and the idler photon $r'$. It decomposes as 
\begin{align}
\psi(r,r')&=R_1(r)e^{iS_1(r)}
R_1'(r')e^{iS_1'(r')}\nonumber\\
&+R_2(r)e^{iS_2(r)} R_2'(r')e^{iS_2'(r')}.
\end{align} Again, primed variables correspond to the idler photon. For the final amplitude $R$ and the phase $S$ we find
\begin{equation}
R^2 =(R_1R_1')^2 + (R_2R_2')^2 + 2R_1R_1'R_2R_2'\cos\Delta \phi,
\end{equation} by the law of cosines, where $\Delta\phi=(S_2+S_2')-(S_1+S_1')$. Also, 
\begin{equation}
\tan S = \frac{R_1R_1'\sin(S_1+S_1')+R_2R_2'\sin(S_2+S_2')}{R_1R_1'\cos(S_1+S_1')+R_2R_2'\cos(S_2+S_2')}.
\end{equation}
We need to evaluate this term for each trajectory. For the photon travelling trough the upper slit the entangled pair is created at this slit, and since the probability of creating an entangled pair at the lower slit is zero when the photon does not pass through it, $R_2'=0$ (since $R_2'$ has no support in the upper slit). Importantly, $R_2\neq 0$ at points where $R_1$ has support. Having said that, vanishing $R_2'$ on this trajectory cancels out overlapping terms, so that $R^2=(R_1R_1')^2$ and interference in the quantum potential vanishes. Recall that the quantum potential is evaluated at the positions of all the particles involved. Likewise, if the photon's path goes through the lower slit, $R_1'=0$. Thus, $R^2=(R_2R_2')^2$ and interference vanishes as before. The guiding phase in the former case yields
\begin{equation}
S= S_1(r)+S_1'(r').
\end{equation} That means that the guidance equation for the signal photon becomes independent of $S_2$ and $S_2'$:

\begin{equation}
p_1=\nabla_{r}S= \nabla_{r}S_1(r),
\end{equation} with $p_1$ the particle's momentum.\footnote{Again, we should talk about massive particles for the guidance equation to make sense. However, the results for photons are equal.} The idler photon then continues to travel to detector $D_4$ or $D_1$. Similarly, in the latter case the signal photon is independent of $S_1$ and $S_1'$. The idler photon then continues to travel to detector $D_3$ or $D_2$. The gradients $\nabla S_1, \ \nabla S_2$ (and consequently the momentum) point in the radial direction away from the slits. All we need to know is that a definite result has actually occurred (such as `the signal photon has passed the upper slit', or `the idler photon follows a path towards detector $D_4$'). Then, all of the other potential states give no contribution to the guidance equation so that the interference term cancels.

I will now turn to the situation where the quantum eraser is present, but we remove the two beamsplitters reflecting the idler photons into the which-path detectors. The question is whether the trajectories change when we consider the quantum potential of the eraser.
Recall the wavefunction of the system when the idler photon has passed the eraser:
\begin{align}
\label{eraser}
\psi &= \frac{1}{2}(\psi_1\otimes (i\psi_{D1}-\psi_{D2})+\psi_2\otimes (-\psi_{D1}+i\psi_{D2})) \nonumber \\
&= \frac{1}{2}((i\psi_{1}-\psi_{2})\otimes\psi_{D1} + (-\psi_{1}+i\psi_{2})\otimes\psi_{D2}).
\end{align}

\noindent Or in polar form
\begin{align}
\label{bohmeraser}
\psi&= R_1(r)e^{iS_1(r)}(R_{D_1}(r')e^{iS_{D_1}(r')+i\frac{\pi}{2}}\\
&-R_{D_2}(r')e^{iS_{D_2}(r')})\nonumber \\
&+ R_2(r)e^{iS_2(r)}(-R_{D_1}(r')e^{iS_{D_1}(r')}\nonumber\\
&+R_{D_2}(r')e^{iS_{D_2}(r')+i\frac{\pi}{2}})\nonumber.
\end{align} Consequently, unlike in the case without the eraser, here the signal photon is guided by a potential with contributions both from $R_1$ and $R_2$. Indeed, assume the idler photon to end in the path leading to detector $D_1$. That means $R_{D_2}=0$ and the trajectory of the signal photon is determined by
\begin{align}
& R_1(r)e^{iS_1(r)}R_{D_1}(r')e^{iS_{D_1}(r')+i\frac{\pi}{2}} \nonumber \\
-& R_2(r)e^{iS_2(r)}R_{D_1}(r')e^{iS_{D_1}(r')},
\end{align} and vice versa by
\begin{align}
&- R_1(r)e^{iS_1(r)}R_{D_2}(r')e^{iS_{D_2}(r')} \nonumber
\\ &+R_2(r)e^{iS_2(r)}R_{D_2}(r')e^{iS_{D_2}(r')+i\frac{\pi}{2}}
\end{align} if the idler photon travels toward detector $D_2$. In both cases the paths are those wiggly trajectories  which photons take in the usual double slit experiment (up to a phase shift). These trajectories produce the same interference patterns that we came across in Figure \ref{fig:patterns}. Bear in mind that if added, they produce a clump pattern.    

\begin{figure}[H]
 		\centering
 		\begin{subfigure}[b]{0.3\textwidth}
 	  			\includegraphics[width=\textwidth]{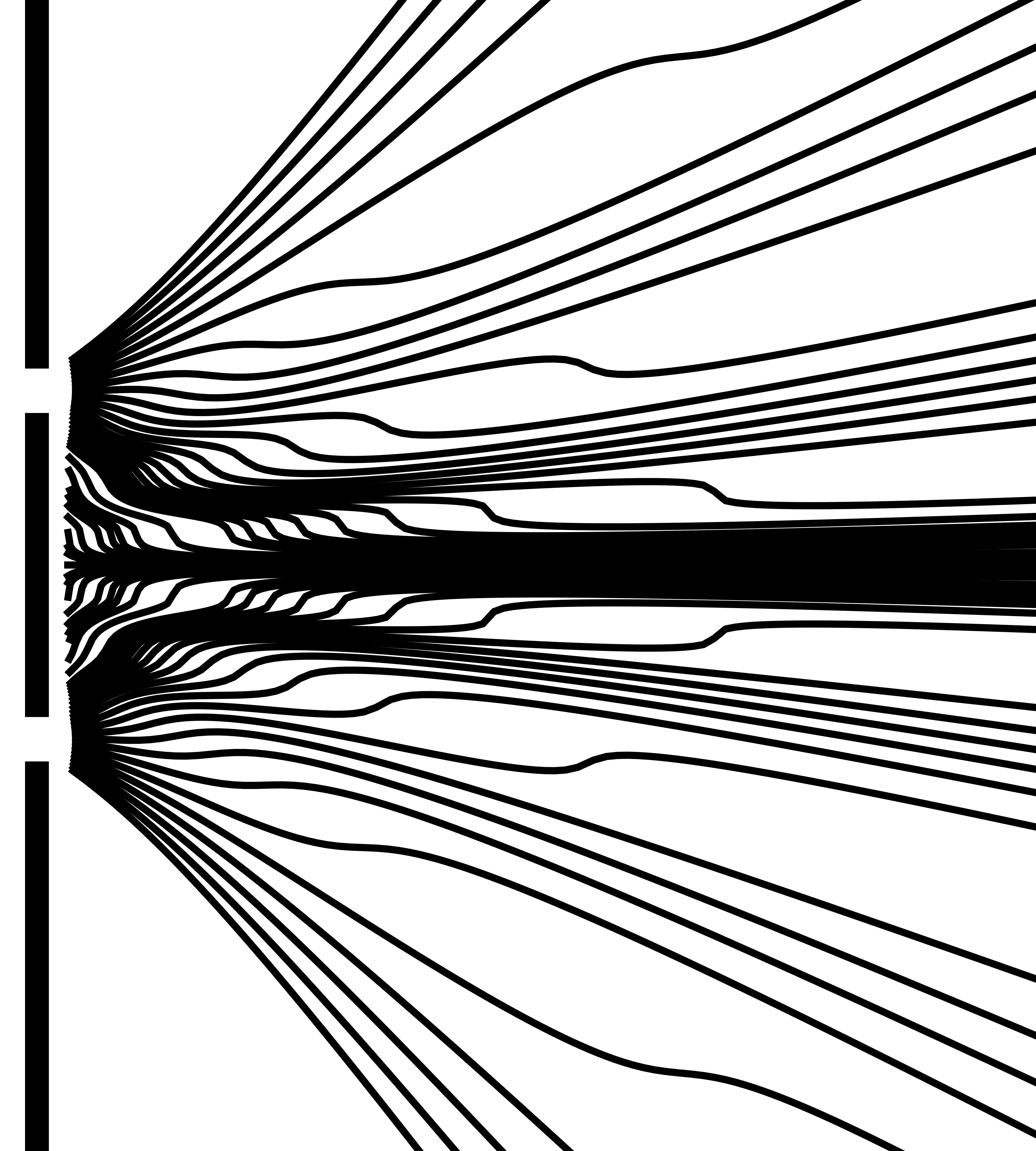}
   				\caption{}
   				\label{fig:s1}
 		\end{subfigure}
 		~
 		\begin{subfigure}[b]{0.3\textwidth}
   				\includegraphics[width=\textwidth]{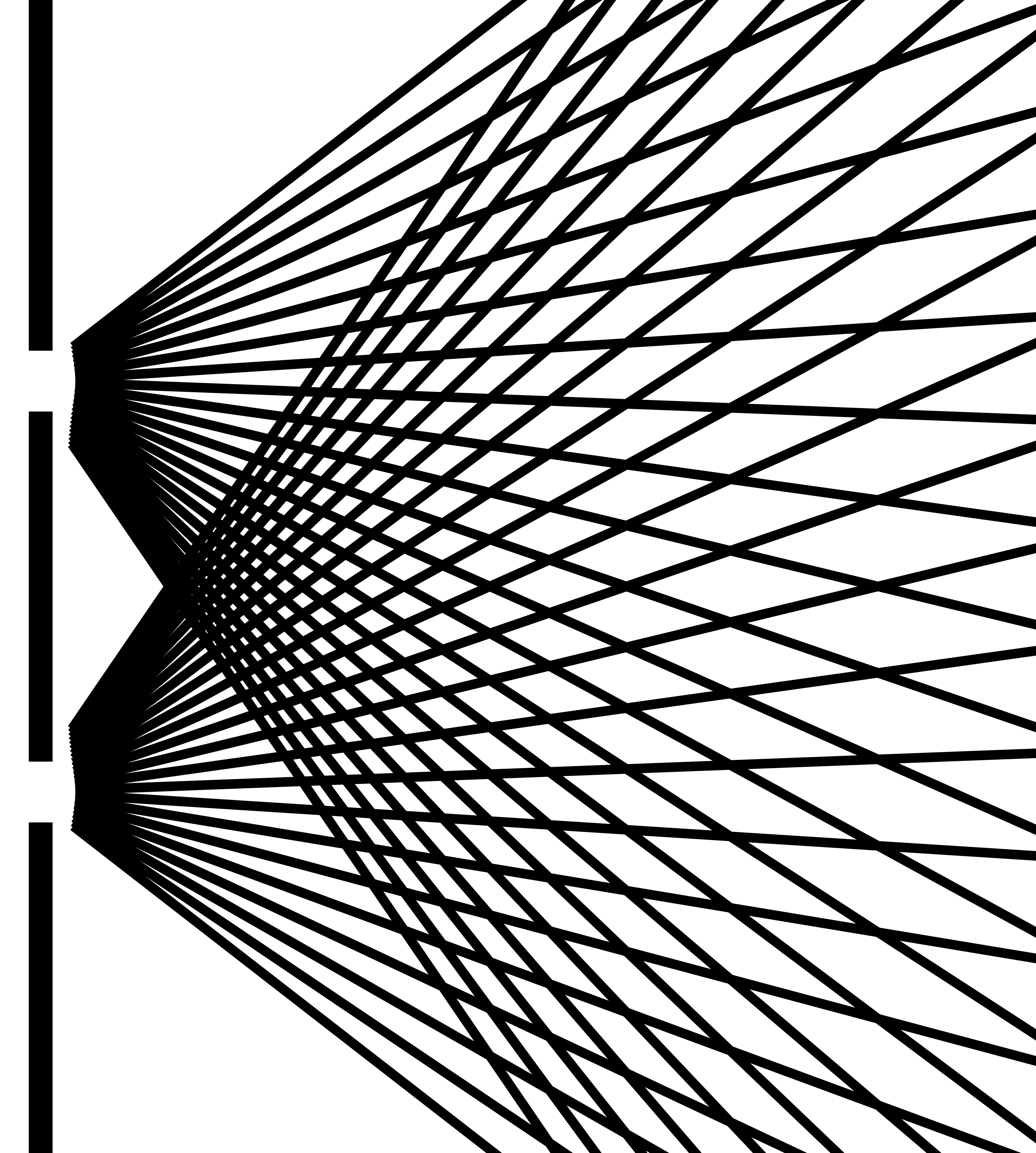}
   				\caption{}
   				\label{fig:s2}
 		\end{subfigure}
		~
 		\begin{subfigure}[b]{0.3\textwidth}
   				\includegraphics[width=\textwidth]{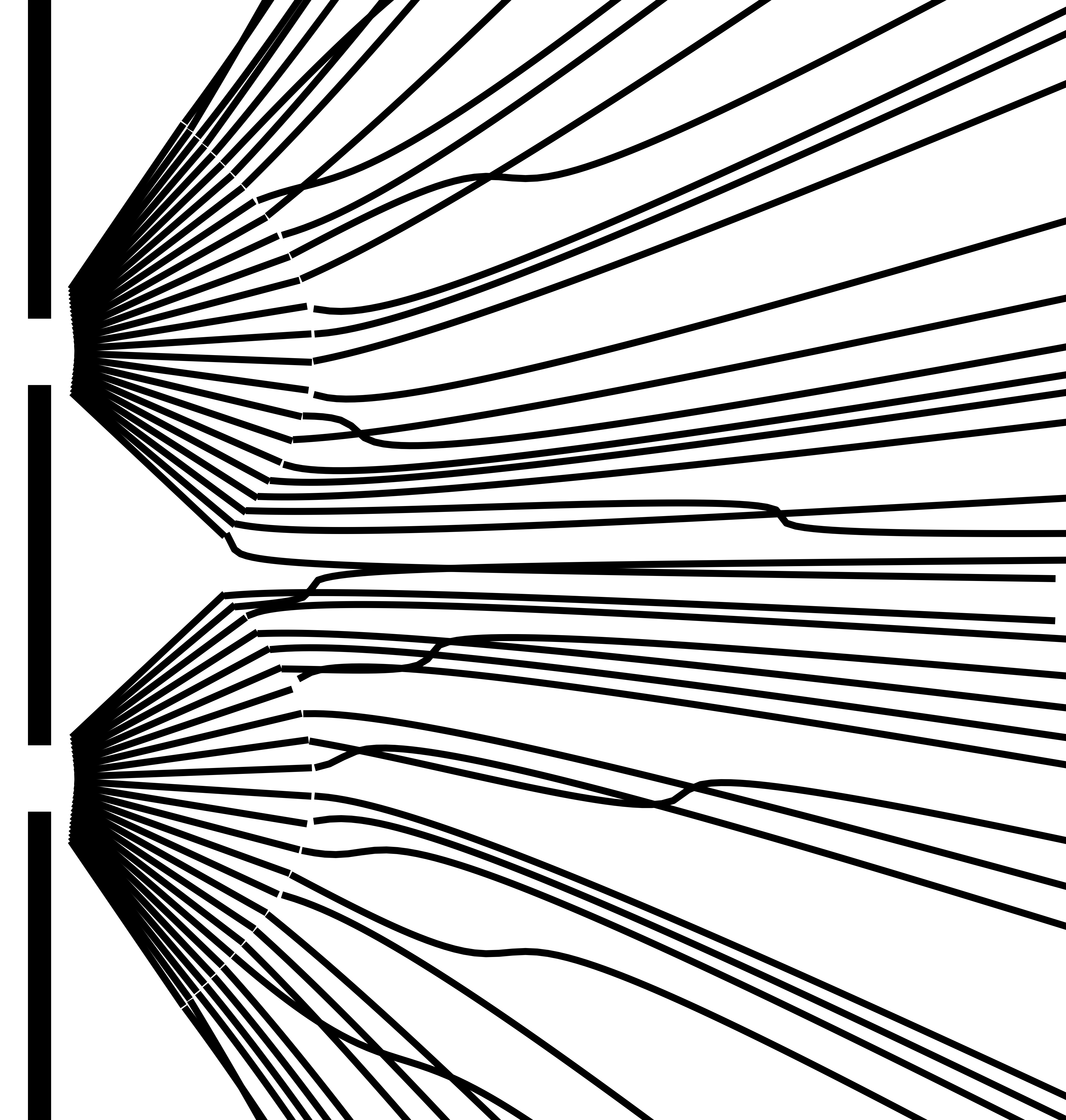}
   				\caption{}
   				\label{fig:s3}
 		\end{subfigure}

		\caption{The signal photon follows different trajectories depending on when the idler photon encounters the quantum eraser. \textbf{(a)} The well-known wiggly trajectories that lead to an interference pattern in a usual double slit experiment. \textbf{(b)} In the case where the idler photon hits the quantum eraser after the signal photon arrives at the screen (which is how the experiment is set up in \cite{Kim1999}), the signal photon moves on straight lines.  \textbf{(c)} Before the idler photon has encountered the quantum eraser the signal photon follows straight lines. When the idler photon travels to detector $D_1$ or $D_2$, a jump in the guidance relation happens, leading to trajectories as in the interfering case.}
		\label{fig:Doppelspalt}	
\end{figure}

The eraser drastically changes the wavefunction, but at the same time the signal photon's past trajectory is not influenced by the change. Depending on when the idler photon enters the region between eraser beamsplitter and detectors $D_1$ or $D_2$, the signal photon jumps from moving on straight lines to following wavy trajectories typical for interference. This is striking, for the effects on the signal photon are mediated superluminally, in conflict with special relativity. On the other hand, this should not be surprising, for non-locality is one of the features of a hidden variable theory like de Broglie-Bohm's. In the experiment of \cite{Kim1999} the moment in time when the idler photon encounters the eraser is always after the signal photon hits the detector. Therefore, the Bohmian trajectories in that case look like the straight lines in Figure \ref{fig:s2}. This shows that it is possible two observe interference effects without wavy trajectories in de Broglie-Bohm theory. That is, after the post-selection of the correlated subensembles of signal and idler photon, the two phase shifted interference patterns are recovered (even without the guidance of a superposition state!). If one adjusted the delay and shorten the optical length of the idler photon such that it passes through the eraser during the signal photon travelling toward $D_0$, the trajectories would look like those in Figure \ref{fig:s3}.

Let us recap. There are two ways in which interference fringes can emerge at the detector $D_0$. When the idler photon arrives at the eraser during the flight of the signal photon, then the signal photon continues to move on wiggly lines giving rise to fringes. There is no change of the past whatsoever. When the idler photon arrives after the signal photon encounters $D_0$, the trajectories are straight lines (see Figure \ref{fig:Doppelspalt}). In this case, selecting out interference patterns by conditioning on $D_1$ and $D_2$ does not change trajectories of the past. The reason we can extract interference fringes is that one subset of the trajectories of the signal photon is consistent with the idler photon being detected at $D_1$ (interference fringes), and another subset is consistent with a detection in $D_2$ (anti-fringes), and both add up to a clump pattern. This is the case in the experiments by Kim \textit{et al.} and causes confusion if we do not consider conditioning on the signal photon, thus calling for the need of `backwards in time influence' to restore the interference outcomes. It also trivially follows from my analysis that there is no need to invoke `entanglement in time'. For I make no use of any non-standard features of standard quantum mechanics or de Broglie-Bohm theory. Pilot wave dynamics restores the conventional view of the world as particles having a definite trajectory and past. In Wheeler's view the past comes into existence only after the measurement in the present, but my analysis gives an account that consistently attributes a past to the photon's trajectory.

\section{Conclusion}

The delayed choice quantum eraser experiment resembles a Bell-type experiment and thus is not more mysterious than that. 
There is no need to invoke a notion such as `the present action determines the past'. I have shown this to be fairly straightforward in the Bell framework. The original puzzle arises due to the symmetry property that the time ordering which party measures first is irrelevant for the statistics of the outcomes and this allows for an alternative explanation in terms of `action into the past'. But such is unwarranted. One can treat the same phenomenon in two different ways given this property, but one of them  looks as if something would have to change in the past.  

We can consistently derive the probabilities for different measurement outcomes in the delayed choice quantum eraser experiment from standard quantum mechanics. When the idler photon is manipulated in a way that provides which-path information about the signal photon, detector $D_0$ does not show interference, even if conditioned on the idler photon's specific measurement results. On the other hand, if the idler photon is detected such that the measurement irrevocably erases  which-path information about the signal photon, then too the interference patterns reappear. Those distributions are complementary in the sense that they add up to a clump pattern. Further, only conditioned on the detector outcomes of the idler photon can the patterns be extracted.

I have shown that both in standard quantum mechanics as well as in the de Broglie-Bohm theory the experiment can be understood without invoking `backwards in time influence'. Properly conditioning the state of the system without neglecting the measurement on the signal photons explains why there is no paradox. The seemingly retroactive action disappears if the effects of measurement on the state of the signal photon is considered to also change the overall state. In the de Broglie-Bohm theory the particle takes one definite trajectory and during its motion does not change its past. However, the idler photon may determine the pilot wavefunction of the signal photon depending on when the idler photon passes the quantum eraser. Most importantly, de Broglie-Bohm theory allows one to consistently construct the trajectories the photons have taken in the past. 

Among the double slit delayed choice experiment there are further cases like delayed choice entanglement swapping or delayed choice Bell experiments. Are these experiments all of the same kind? I presume they all can be elucidated in a similar fashion as I did in this paper. Considerations on delayed choice experiments could also have consequences on ideas like entanglement realism. What is the status of entanglement if quantum effects like interference, entanglement swapping, and violation of Bell inequalities can equally well be confirmed via post selection in delayed choice scenarios? Questions of this kind are ripe for investigation. 

\section*{Acknowledgments}

I thank Christopher Timpson for fruitful discussions and three reviewers for helpful comments.

\bibliography{library}

\begin{thebibliography}{}

\bibitem[Aharonov and Zubairy, 2005]{aharonov2005time}
Aharonov, Y. and Zubairy, M.~S. (2005).
\newblock {Time and the quantum: erasing the past and impacting the future}.
\newblock {\em Science}, 307(5711):875--879.

\bibitem[Bohm and Hiley, 2006]{bohm2006undivided}
Bohm, D. and Hiley, B.~J. (2006).
\newblock {\em {The undivided universe: An ontological interpretation of
  quantum theory}}.
\newblock Routledge.

\bibitem[Bohr, 1961]{bohr1961atomic}
Bohr, N. (1961).
\newblock {\em {Atomic physics and human knowledge}}.
\newblock Science Editions New York.

\bibitem[Bohr, 1996]{bohr1996discussion}
Bohr, N. (1996).
\newblock {Discussion with Einstein on epistemological problems in atomic
  physics}.
\newblock In {\em Niels Bohr Collected Works}, volume~7, pages 339--381.
  Elsevier.

\bibitem[Egg, 2013]{Egg2013}
Egg, M. (2013).
\newblock {Delayed-Choice Experiments and the Metaphysics of Entanglement}.

\bibitem[Eichmann et~al., 1993]{eichmann1993young}
Eichmann, U., Bergquist, J.~C., Bollinger, J.~J., Gilligan, J.~M., Itano,
  W.~M., Wineland, D.~J., and Raizen, M.~G. (1993).
\newblock {Young's interference experiment with light scattered from two
  atoms}.
\newblock {\em Physical review letters}, 70(16):2359.

\bibitem[Einstein et~al., 1935]{PhysRev.47.777}
Einstein, A., Podolsky, B., and Rosen, N. (1935).
\newblock {Can Quantum-Mechanical Description of Physical Reality Be Considered
  Complete?}
\newblock {\em Phys. Rev.}, 47(10):777--780.

\bibitem[Ellerman, 2011]{ellerman2011very}
Ellerman, D. (2011).
\newblock {A Very Common Fallacy in Quantum Mechanics: Superposition, Delayed
  Choice, Quantum Erasers, Retrocausality, and All That}.
\newblock {\em arXiv preprint arXiv:1112.4522}.

\bibitem[Ellerman, 2015]{Ellerman2015}
Ellerman, D. (2015).
\newblock Why delayed choice experiments do not imply retrocausality.
\newblock {\em Quantum Studies: Mathematics and Foundations}, 2(2):183--199.

\bibitem[Englert and Bergou, 2000]{englert2000quantitative}
Englert, B.-G. and Bergou, J.~A. (2000).
\newblock {Quantitative quantum erasure}.
\newblock {\em Optics communications}, 179(1):337--355.

\bibitem[Englert et~al., 1999]{doi:10.1119/1.19257}
Englert, B.-G., Scully, M.~O., and Walther, H. (1999).
\newblock {Quantum erasure in double-slit interferometers with which-way
  detectors}.
\newblock {\em American Journal of Physics}, 67(4):325--329.

\bibitem[Gaasbeek, 2010]{Gaasbeek2010}
Gaasbeek, B. (2010).
\newblock {Demystifying the Delayed Choice Experiments}.
\newblock {\em Arxiv}, page~7.

\bibitem[Greene, 2004]{brian2004fabric}
Greene, B. (2004).
\newblock The fabric of the cosmos.
\newblock {\em Space, Time and the Texture of Reality (Albert A. Knopf, New
  York, 2004)}.

\bibitem[Healey, 2012]{healey2012quantum}
Healey, R. (2012).
\newblock {Quantum theory: a pragmatist approach}.
\newblock {\em The British Journal for the Philosophy of Science},
  63(4):729--771.

\bibitem[Hiley and Callaghan, 2006]{hiley2006erased}
Hiley, B.~J. and Callaghan, R.~E. (2006).
\newblock {What is erased in the quantum erasure?}
\newblock {\em Foundations of Physics}, 36(12):1869--1883.

\bibitem[Kastner, 2019]{kastner2019delayed}
Kastner, R. (2019).
\newblock The ‘delayed choice quantum eraser’neither erases nor delays.
\newblock {\em Foundations of Physics}, pages 1--11.

\bibitem[Kim et~al., 1999]{Kim1999}
Kim, Y.-H., Yu, R., Kulik, S.~P., Shih, Y.~H., and Scully, M. .~O. (1999).
\newblock {A Delayed Choice Quantum Eraser}.
\newblock pages 1--4.

\bibitem[Kwiat and Englert, 2004]{kwiat2004science}
Kwiat, P.~G. and Englert, B.~G. (2004).
\newblock {Science and ultimate reality: quantum theory, cosmology and
  complexity}.

\bibitem[Mohrhoff, 1999]{doi:10.1119/1.19258}
Mohrhoff, U. (1999).
\newblock {Objectivity, retrocausation, and the experiment of Englert, Scully,
  and Walther}.
\newblock {\em American Journal of Physics}, 67(4):330--335.

\bibitem[Peres, 2000]{Peres2000}
Peres, A. (2000).
\newblock {Delayed choice for entanglement swapping}.
\newblock {\em Journal of Modern Optics}, 47(2-3):139--143.

\bibitem[Scully and Dr{\"{u}}hl, 1982]{scully1982quantum}
Scully, M.~O. and Dr{\"{u}}hl, K. (1982).
\newblock {Quantum eraser: A proposed photon correlation experiment concerning
  observation and" delayed choice" in quantum mechanics}.
\newblock {\em Physical Review A}, 25(4):2208.

\bibitem[Shah and Qureshi, 2017]{shah2017quantum}
Shah, N.~A. and Qureshi, T. (2017).
\newblock Quantum eraser for three-slit interference.
\newblock {\em Pramana}, 89(6):80.

\bibitem[von Weizs{\"{a}}cker, 1941]{von1941deutung}
von Weizs{\"{a}}cker, C.~F. (1941).
\newblock {Zur deutung der Quantenmechanik}.
\newblock {\em Zeitschrift f{\"{u}}r Physik}, 118(7-8):489--509.

\bibitem[Walborn et~al., 2002]{walborn2002double}
Walborn, S.~P., Cunha, M. O.~T., P{\'{a}}dua, S., and Monken, C.~H. (2002).
\newblock {Double-slit quantum eraser}.
\newblock {\em Physical Review A}, 65(3):33818.

\bibitem[Wheeler, 1978]{wheeler1978past}
Wheeler, J.~A. (1978).
\newblock {The “past” and the “delayed-choice” double-slit experiment}.

\end{thebibliography}
\end{document}